\documentclass[useAMS,usenatbib]{mn2e}

\usepackage{natbib}
\usepackage{graphicx}
\usepackage{multirow}
\usepackage{subfig}
\usepackage{amsmath}
\usepackage{amssymb}
\usepackage{epstopdf}
\newcommand{\araa}{ARA\&A}
\newcommand{\apj}{ApJ}
\newcommand{\apjs}{ApJS}
\newcommand{\apjl}{ApJL}
\newcommand{\aj}{AJ}
\newcommand{\mnras}{MNRAS}
\newcommand{\aap}{A\&A}
\newcommand{\procspie}{SPIE}
\newcommand{\pasj}{PASJ}
\newcommand{\pasp}{PASP}

\title[High angular resolution mid-IR emission in UGC~5101]{A deep look at the nuclear region of UGC~5101 through high angular resolution mid-IR data with GTC/CanariCam}

\author[Mart\'{\i}nez-Paredes et al.]
{\parbox{\textwidth}{M. Mart\'{\i}nez-Paredes,$^{1}$\thanks{E-mail:
      mariela@inaoep.mx} 
    A. Alonso-Herrero,$^{2}$\thanks{Visiting Professor, Department of Physics and Astronomy,
University of Texas at San Antonio}
I. Aretxaga,$^{1}$   
  C. Ramos Almeida,$^{3,4}$\thanks{Marie Curie Fellow.} 
   A. Hern\'an-Caballero,$^2$ O. 
Gonz\'alez-Mart\'{\i}n,$^{3,4,5}$  M. Pereira-Santaella,$^{6}$
C. Packham,$^{7}$ A. Asensio Ramos,$^{3,4}$   T. D\'{\i}az-Santos,$^{13,14}$ M. Elitzur,$^{15,16}$ P. Esquej,$^{8}$ I. Garc\'{\i}a-Bernete,$^{3,4}$ M. Imanishi,$^{10,11,12}$ N. A. Levenson,$^{9}$ J. M. Rodr\'{\i}guez Espinosa$^{3,4}$} 
\vspace{0.4cm} \\
$^{1}$Instituto Nacional de Astrof\'{\i}sica, \'Optica y Electr\'onica (INAOE), Luis Enrique Erro 1, Sta. Ma. Tonantzintla, Puebla, Mexico\\
$^{2}$Instituto de F\'{\i}sica de Cantabria, CSIC-UC, E-39005 Santander,
Spain\\
$^3$Instituto de Astrof\'{\i}sica de Canarias (IAC), E-38205 La Laguna,
Tenerife, Spain\\
$^4$Departamento de Astrof\'{\i}sica, Universidad de la Laguna (ULL),
E-38206 La Laguna, Tenerife, Spain\\
$^{5}$Centro de Radioastronom\'{\i}a y Astrof\'{\i}sica UNAM, Apartado
Postal 3-72 (Xangari), 58089 Morelia, Michoac\'an, Mexico\\
$^{6}$Centro de Astrobiolog\'{\i}a, CSIC-INTA, E-28850 Torrej\'on de
Ardoz, Madrid, Spain\\
$^{7}$Department of Physics and Astronomy,
University of Texas at San Antonio, San Antonio, TX 78249, USA\\
$^{8}$Departamento de Astrof\'{\i}sica, Universidad Complutense de
Madrid, E-28040 Madrid, Spain\\ 
$^{9}$Gemini Observatory, Casilla 603, La Serena, Chile\\
$^{10}$Subaru Telescope, 650 North A'ohoku Place, Hilo, HI 96720, USA\\
$^{11}$National Astronomical Observatory of Japan, 2-21-1 Osawa, Mitaka,
Tokyo 181-8588, Japan\\
$^{12}$Department of Astronomical Science,
The Graduate University for Advanced Studies (SOKENDAI), 2-21-1 Osawa,
Mitaka, \\
Tokyo 181-8588, Japan\\
$^{13}$Spitzer Science Center, California Institute of Technology, MS 220-6,
Pasadena, CA 91125, USA\\
$^{14}$Nucleo de Astronom\'{\i}a de la Facultad de Ingenier\'{\i}a,
Universidad Diego 
Portales, Av. Ej\'ercito Libertador 441, Santiago, Chile\\
$^{15}$Astronomy Department, University of California, Berkeley,  CA
94720-3411, USA\\
$^{16}$Physics and Astronomy, University of Kentucky, Lexington, KY
40506-0055, USA\\
}

\begin{document}

\date{Accepted ---. Received ---; in original form ---}

\pagerange{\pageref{firstpage}--\pageref{lastpage}} \pubyear{2014}

\maketitle

\label{firstpage}

\begin{abstract}

We present an analysis of the nuclear infrared (IR, 1.6 to 18 $\mu$m) emission of the ultraluminous IR galaxy UGC~5101 to derive the properties of its active galactic nucleus (AGN) and its obscuring material. We use new mid-IR high angular resolution ($0.3-0.5$ arcsec) imaging using the Si-2 filter ($\lambda_{C}=8.7\, \mu$m) and $7.5-13$ $\mu$m spectroscopy
taken with CanariCam (CC) on the 10.4m Gran Telescopio CANARIAS. We also use archival HST/NICMOS and Subaru/COMICS imaging and {\it Spitzer}/IRS spectroscopy. We estimate the near- and mid-IR unresolved nuclear emission by modelling the imaging data with {\sc GALFIT}. We decompose the {\it Spitzer}/IRS and CC spectra using a power-law component, which represents the emission due to dust heated by the AGN, and a starburst component, both affected by foreground extinction. We model the resulting unresolved near- and mid-IR, and the starburst subtracted CC spectrum with the {\sc CLUMPY} torus models of Nenkova et al. The derived geometrical properties of the torus, including the large covering factor and the high foreground extinction needed to reproduce the deep $9.7\, \mu$m silicate feature, are consistent with the lack of strong AGN signatures in the optical. We derive an AGN bolometric luminosity $L_{bol}\sim1.9\times10^{45}\,$erg s$^{-1}$ that is in good agreement with other estimates in the literature.

\end{abstract}


\section{Introduction}

 Ultraluminous infrared galaxies (ULIRGs) are among the most luminous
 objects in the local universe, with both their luminosities
 ($L_{\rm IR}>10^{12}L_{\odot}$ emerging mainly in the far-infrared) and space
 densities  being similar to those of quasars (e.g. Sanders \& Mirabel
 1996). ULIRGs have  large molecular gas concentrations in their
 central kpc regions (e.g. Downes \& Solomon 1998) with densities
 comparable to  those of the stars of elliptical galaxies. For increasing
 infrared (IR) luminosities, their mid-IR slopes and silicate depths also
 increase, suggesting  that their nuclei are more obscured and more compact
 (Stierwalt et al. 2013 and references therein). Kormendy \& Sanders
 (1992)  proposed 
 that ULIRGs evolve into elliptical galaxies through merger-induced
 dissipative collapse. In this scenario, these mergers first go
 through a luminous starburst phase, followed by a dust-enshrouded
 active galactic nucleus (AGN)
 phase, and finally evolve into  an optically bright, {\it naked} quasar
 once they either consume or shed their shells of gas and dust
 (Sanders et al. 1988a). 

The target of interest in this paper is the nearby ($z=0.039$; Kim 1995)
ULIRG UGC~5101 (IRAS~09320+6134) with a luminosity distance
$d=168\,$Mpc  
  (1 arcsec $=755$ pc) for
$H_{0}=73$ km s$^{-1}$Mpc$^{-1}$, $\Omega_{M}=0.27$,
$\Omega_{\Lambda}=0.73,$ and an IR luminosity $L_{\rm
  IR}=10^{12}\,$L$_{\odot}$ (Sanders et 
al. 1988a). In the optical and near-IR the galaxy 
has a single, very red nucleus within a disturbed
morphology, suggestive of a recent interaction and merger (Sanders et
al. 1988b; Surace et al. 2000; Scoville et al. 2000). 

Optically the AGN has been classified as a 
Seyfert 1.5 (Sanders et al. 1988a),
a LINER (Veilleux et al. 1995), and a Seyfert 2 (Yuan et al. 2010). Mid-IR spectroscopy taken with {\it 
  ISO} (Genzel et al. 1998)  as well as ground-based $L$-band
spectroscopy (Imanishi et al. 2001)
demonstrated the presence of a powerful nuclear
starburst. Using ground-based
high angular resolution mid-IR imaging, Soifer et al. (2000) 
showed that
$\approx60\%$ of the {\it IRAS} flux at 12 $\mu$m comes from the central
$4\,{\rm arcsec}$ ($\sim 3\,$kpc) and that almost half of this arises in an unresolved core. Based
on its {\it IRAS} colors, UGC~5101 
is classified as a cold, starburst-dominated, far-IR source. However, {\it Chandra} and {\it XMM-Newton} data
detected the Fe-K$\alpha$ line, which together with the low observed
$L_{\rm 2-10 keV}/L_{\rm IR}$   and $L_{\rm 2-10
keV}/L_{\rm [OIII]}$ ratios likely indicate the presence of a
Compton-thick AGN (Ptak
et al. 2003; Imanishi et 
al. 2003;  Gonz\'alez-Mart\'{\i}n et al. 2009). At radio wavelengths, the nucleus
of UGC~5101 has a high
brightness temperature ($T>10^7\,$K) with a morphology
resembling that of a jet, again pointing to the
presence of an AGN (Lonsdale 
et al. 1995, 2003). Using {\it Spitzer} IR spectrograph (IRS, Houck et
al. 2004) data, Armus et
al. (2004, 2007) detected the mid-IR fine structure line of [Ne\,{\sc v}] at
$14.3\,\mu$m, which confirms the presence of an AGN in the 
  nucleus of this galaxy. 

\begin{figure*}
\center
\vspace{-2cm}
\includegraphics[scale=0.65,angle=90]{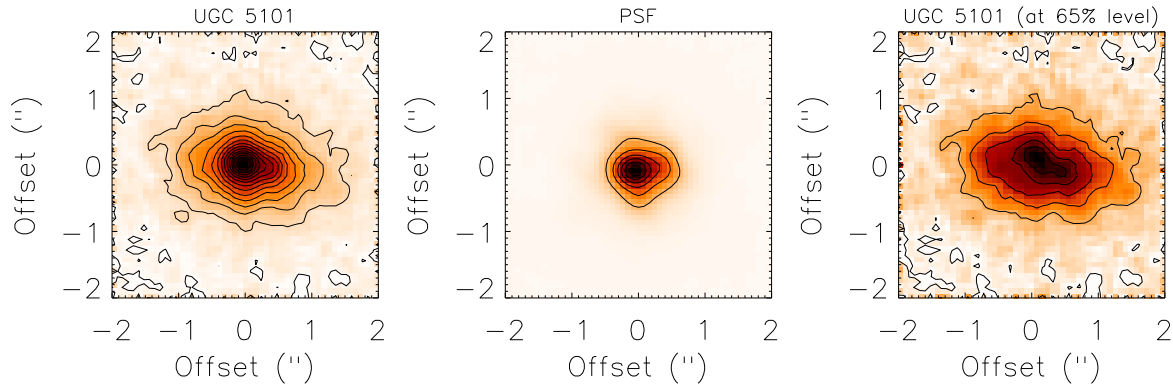}
\vspace{-2cm}
\caption{The left panel is the fully reduced
      GTC/CC $8.7\,\mu$m (Si-2 filter)
      image of UGC~5101, in the middle is the PSF standard star shown in a
      linear scale. The right panel is
    the PSF-scaled  
    (at the $65$\% peak level, see section 3.1.2)
    subtracted image of UGC~5101. The
  lowest contour is 
  $3\sigma$ over the background, with the next contours in $2\sigma$ steps (except in the PSF image). The orientation of all the
    images is North up, East to the left. The physical scale of the UGC 5101 images is 755 pc/$''$.} 
\label{fig1} 
\end{figure*}

UGC~5101 offers an ideal scenario for studying the interplay between
nuclear activity and star formation in a nearby ULIRG. The 
$\sim5-10$\,arcsec spatial
resolution of {\it Spitzer}/IRS  only  
  allowed to study this interplay in ULIRGs on 
kpc-scales, although with extremely good sensitivity.  For
  instance, Veilleux et al. (2009) used different indicators on the
  {\it Spitzer}/IRS
  spectrum of this galaxy and estimated an average IR luminosity, whereas Armus et al. (2007) estimated a 30\% contribution from a starburst component. 

Here we present new high spatial resolution ($\sim0.3\,{\rm
  arcsec} \sim 230\,$pc)
 mid-IR imaging and spectroscopy of UGC~5101
obtained with CanariCam  (CC, Telesco et al. 2003; Packham et
al. 2005) on  the 10.4\,m Gran 
Telescopio  CANARIAS (GTC). We use these observations to
disentangle the  unresolved  mid-IR emission, which is  presumably
  due to dust heated by the AGN, from that due to star formation in the
  nuclear region of the host
galaxy. By modelling
the unresolved near-IR and mid-IR emission
with the {\sc CLUMPY} torus models of
Nenkova et al. (2008a, b) we aim to derive the
physical and geometrical properties of the obscuring
 material surrounding  the AGN of 
  UGC~5101. The paper is organized as follows. Section~2
  presents the observations and data reduction. In Section~3 we
  estimate the unresolved emission in the near-IR and mid-IR using
  the imaging and spectroscopic data. In Section~4 we model the
  unresolved near and mid-IR emission of UGC~5101 and derive the
  properties of the AGN  and its obscuring
  material. Section~5 gives the conclusions of this work.

\section{Observations and data reduction}
\subsection{GTC/CC mid-IR imaging and spectroscopy}
As part
of an ESO/GTC large programme (ID: 182.B-2005, PI:
A. Alonso-Herrero) to observe a large sample of local AGN, in January 6th, 2014, we observed in queue mode UGC~5101 with CC
on the GTC in El Roque de los Muchachos Observatory (La Palma,
Spain). CC uses a Raytheon $320\times240$  Si:As 
detector that covers a field of view (FOV) of $26\,{\rm
  arcsec}\times19\,{\rm arcsec}$ on the 
sky with a plate scale of  $0.0798\,{\rm arcsec}$ pixel$^{-1}$. We
obtained an imaging observation with the Si-2 filter
($\lambda_{c}=8.7$ 
$\mu$m and width 
$\Delta\lambda_{\rm cut}=1.1$ $\mu$m at $50\%$ cut-on/off) using
  the standard 
mid-IR chop-nod technique. The chop and nod throws were
$15\,{\rm arcsec}$. The airmass 
during the observation was $\sim1.2$. We observed the mid-IR
spectro-photometric standard star HD~79354 (Cohen 1999) before the science target. Allowing for overheads, this was 20 min before the science observations. The standard star is used to perform the photometric
calibration, estimate the full width half
maximum (FWHM) of the science observation, and
perform the point spread function (PSF) subtraction. We  measured the FWHM
of the data by fitting the radial profile of  the standard star with a
Gaussian function and obtained a value of $0.4\,{\rm arcsec}$. Table~\ref{tab0} summarizes the details of the observations.

We also obtained long-slit mid-IR spectroscopy of the nuclear region
of UGC~5101 on the same night. We  used
the low spectral resolution 10 $\mu$m grating, which covers the
$N$-band $\sim7.5-13$ $\mu$m with a nominal spectral resolution of
$R=\lambda/\Delta\lambda\sim175$, and a $0.52\,{\rm arcsec}$ width slit. In
  Table~\ref{tab0} we list the position angles (PA) of the slit,
measured from the North to the 
East. The observing sequence was as follows. We first took
an acquisition image of the galaxy with the Si-2 filter, then centered the slit on the peak mid-IR brightness, and finally
integrated for the on-source  integration times given in Table~\ref{tab0}. 
 We used the same chop-nod parameters as for
the imaging observations. We also observed standard stars using the
same observing sequence to provide the photometric calibration, the
telluric correction, and the slit loss correction. We also used the acquisition image of the standard star to obtain the FWHM of the
spectroscopic observation (see Table \ref{tab0}).

\begin{table}
\caption{Summary of the GTC/CC observations.} 
\begin{minipage}{1.\textwidth}
\begin{tabular}{c|cc} 
UGC 5101 & Imaging & Spectroscopy \\
\hline
Date & 2014.01.06 & 2014.01.06 \\
$NOB\times t_{\rm on}$ (s) & $1\times1224$ & $1\times1242$ \\
Standard & HD~79354 & HD~79354 \\
$NOB\times t_{\rm on-STD}$ (s) &   $1\times60$& $1\times60$ \\
FWHM (arcsec) & 0.4 & 0.3 \\
PA (degree)& 0 & 90 \\
\hline 
\end{tabular} 
\label{tab0}
\end{minipage}
Notes.--- N$_{OB}$ is the number of repetitions performed, $t_{\rm on}$ is the on-source integration time of UGC 5101 and $t_{\rm on-STD}$ is the on-source integration time of the standard star. The FWHM is measured from the image of the standard
star. 
The PA value for imaging is the orientation of the
detector on the sky, while for spectroscopy it indicates the
orientation of the slit.
\end{table}

We reduced the CC data using the {\sc RedCan} pipeline for
the reduction and analysis of ground-based mid-IR CC and T-ReCS 
imaging and spectroscopic data 
(Gonz\'alez-Mart\'in et al. 2013). Briefly, the reduction process of
the imaging data includes sky subtraction, stacking of the individual
images, and rejection of bad images. The flux calibration of the
galaxy image is done using the observation of the standard
star. Fig.~\ref{fig1} (left panel) shows the fully-reduced
$8.7\,\mu$m CC image of 
UGC~5101 together with that of the standard star (middle panel). As also found by 
Soifer et al. (2000) at $12\,\mu$m using the Keck telescope, UGC~5101
is clearly extended in the mid-IR (see 
Section~3.2). 

For the spectroscopy the first three steps of the data reduction are
the same as for the imaging. Additionally, {\sc RedCan} performs the
two-dimensional wavelength calibration of the target and standard star
spectra using sky lines. Then, the trace determination is made using
the observation of the standard star. The last steps of the
  data reduction  include the extraction  of 
the spectra either as
point sources or extended sources and finally, the correction for slit 
losses in the case of point source extractions.  

\subsection{Archival {\it HST}/NICMOS and Subaru/COMICS imaging}
Our target was observed in a single orbit in November 7 of 1997 with the Hubble Space Telescope (HST) using
camera 2 of NICMOS (Thompson et al. 1998) as part of the Guaranteed Time Observations. This camera  uses a $256\times256$
HgCdTe array with 
pixel scales of $0.0762\,{\rm arcsec}$  and $0.0755\,{\rm arcsec}$ pixel$^{-1}$ in $x$ and $y$,
respectively, providing a $\sim19.5\,{\rm arcsec}\times 19.3\,{\rm
  arcsec}$  FOV. We use the images of UGC~5101
taken with the broad-band F160W filter ($\lambda_{\rm c}=1.60\,\mu$m) and
the medium-band  
F222M filter ($\lambda_{\rm c}=2.22\,\mu$m). We
downloaded the five reduced and calibrated images in each of the
 F160W and F222M filters  
from the Mikulski Archive for Space Telescopes (MAST).   We
  combined the individual exposures of each filter using the average of
  the images. These images were
  originally presented and analysed in Scoville et al. (2000).

We also downloaded a fully reduced and calibrated image of UGC~5101
taken in the $Q$-band ($\lambda_{\rm c}=17.7\,\mu$m)  from the German
Virtual Observatory (GAVO) hosted at the Virtual Observatory
(VO). The image was obtained with COMICS on the Subaru
Telescope in Mauna Kea, Hawaii. It has a  FOV and a pixel scale
of $42\,{\rm arcsec}\times32\,{\rm arcsec}$ and $0.13\,{\rm arcsec\,pixel}^{-1}$, respectively. 
For full details on the $Q$-band observations and data reduction we
  refer the reader to Asmus et al. (2014).

\begin{figure*}
\center
\vspace{+1cm}
\includegraphics[angle=90,scale=0.7]{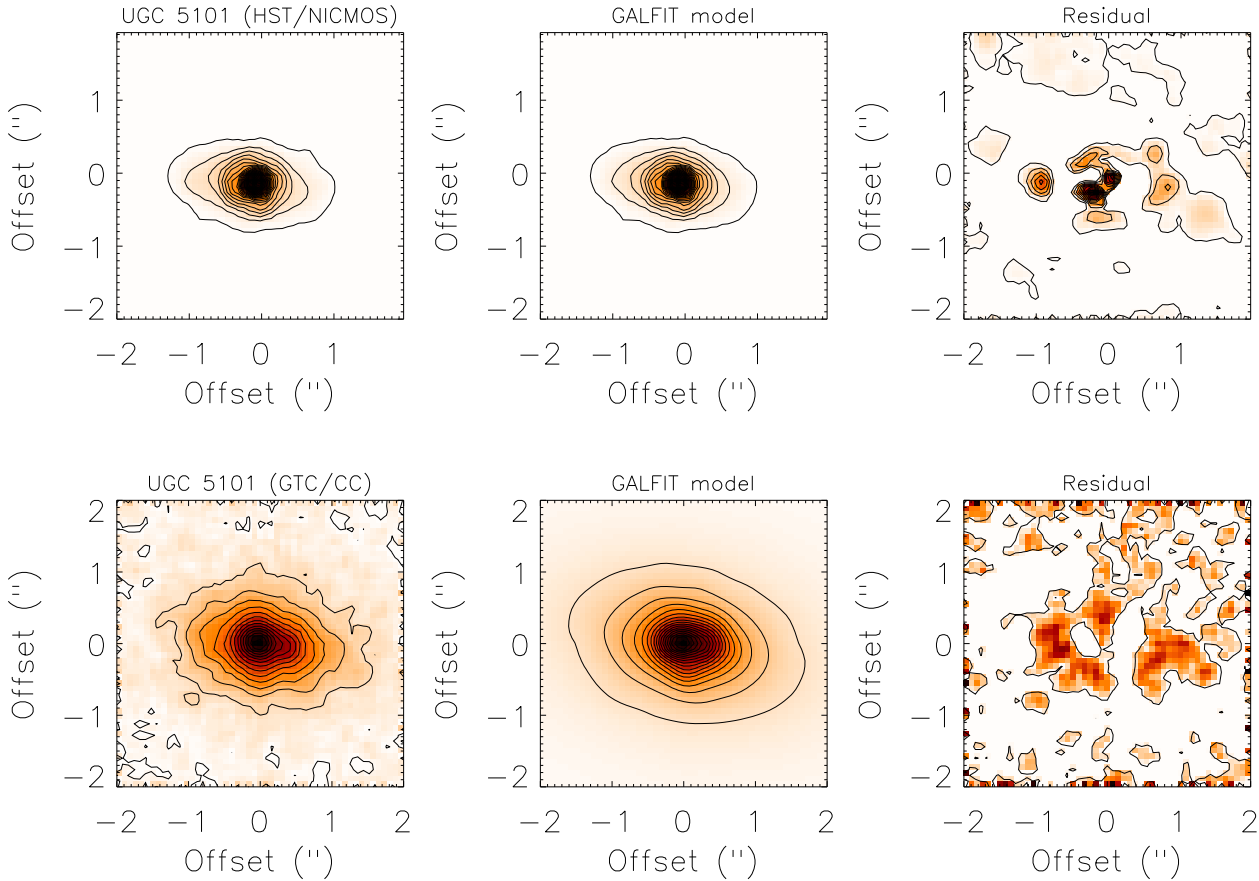}\\ 
\vspace{+1cm}
\caption{\label{fig2}Upper panels: to the left is the combined {\it HST}/NICMOS
  $2.22\,\mu$m image (F222M filter) of the 
  nuclear region of UGC~5101, in the middle the {\sc GALFIT} model image
  (see Table~2 for the parameters), and to the right the residual image after
  subtracting the {\sc GALFIT} model 
  from the original image. The images are shown in
    a linear scale, rotated so that North is up, East is to
    the left. The lowest contour is $3\sigma$ above the background,
    with the next contours in $2\sigma$ steps. Lower panels: same as upper panels but for the reduced GTC/CC $8.7\, \mu$m (Si-2 filter) image. The physical scale of the the images is 755 pc/$''$.}   
\end{figure*}

\subsection{Spitzer/IRS spectroscopy}
UGC~5101 was observed with {\it Spitzer}/IRS using the short-low (SL) and
long-low (LL) modules covering the  following spectral ranges: 
SL1; $7.4-14.5\,\mu$m, SL2; $5.2-7.7\,\mu$m, LL1; and $19.5-38.0\,\mu$m,
LL2; $14.0-21.3\,\mu$m, with a
  spectral resolution of $R\sim60-120$. The SL slit width is
$3.7\,$arcsec,  whereas that of the LL module is $10.5\,$arcsec.
We downloaded the  
fully-calibrated staring mode spectrum from the Cornell Atlas of
{\it Spitzer}/IRS Source (CASSIS v6; Lebouteiller et al. 2011). CASSIS has identified this object as point-like and therefore it uses its {\it optimal extraction} routine to produce the best flux-calibrated spectrum and ensure the best
S/N ratio. We stitched the different module spectra together
by  using the shortest   module SL2 flux as reference spectrum for the
other modules. The scaling factors were 1.1 for LL1
and LL2. 

\section{Analysis}

\subsection{Unresolved nuclear emission from the imaging data}
Even at high spatial resolutions, in the near-IR up to $\lambda\sim2$
$\mu$m, extended stellar emission 
arising in the host galaxy contaminates and even dominates the nuclear
fluxes of AGN, especially in type 2s. At longer wavelengths
$\lambda>3$ $\mu$m contamination by stellar photospheric emission is
greatly  reduced (Alonso-Herrero et al. 2001 and references therein).
Our goal in this section is to 
  use the high angular resolution imaging data to 
  isolate as much as possible the  unresolved nuclear emission
before we attempt to infer the properties of the
obscuring material and the AGN luminosity of 
UGC~5101 (Section~4). To do so, we need to remove any extended
nuclear emission not directly related to the dusty torus and/or
emission arising from dust
heated by young massive stars (see Alonso-Herrero
et al. 2011 and references  therein).

 Although in the past there have been some attempts to measure
the unresolved nuclear emission of UGC~5101 at near- and mid-IR
 wavelengths (e.g. Soifer et al. 2000; Scoville et
al. 2000; Haan et al. 2011; Imanishi et al. 2014; Asmus et al. 2014), the reported values
may still be  contaminated by stellar emission  and emission
due to nuclear star formation.  Therefore, we measure the unresolved emission consistently
in the archival {\it HST}/NICMOS $H$- and $K$-band images and
  COMICS $Q$ band  
images as well as in the new CC image. To do so, we  use
the two-dimensional fitting algorithm {\sc GALFIT} (Peng et al. 2002) to derive both the unresolved nuclear emission and extended emission, as {\sc GALFIT} allows to simultaneously fit an arbitrary
number of components (e.g. PSF, S\'ersic profile, exponential disk,
and/or Gaussian functions) to a galaxy image. It is possible, however, that even at these high angular resolutions ($0.15-0.40''$) there is still some contamination from other components (stellar emission and star formation, see Section 4).

\subsubsection{Near-IR emission}
To model the unresolved nuclear emission we  generated
theoretical Tiny Tim PSFs for the NICMOS camera 2 F160W and F222M
filters. The Tiny Tim PSFs have only three free parameters,
namely the central
coordinates ($x_{c}$ and $y_{c}$) and the total flux. The total
unresolved emission is computed by integrating the flux
over the PSF image assuming that it contains $100\%$ of the light.  
To fit the extended surface brightness distribution, that is, the host
galaxy emission, we  used a
S\'ersic profile (see Peng et al. 2002 for more details on profiles).

\begin{table*} 
\caption{Results from the {\sc GALFIT} modelling using a PSF + S\'ersic
  profile model. }  

\begin{minipage}{1.\textwidth}
\center
\begin{tabular}{lccccccc} 
\hline
Instrument & $\lambda$&$f_{\rm unresol}$ & $f_{\rm Sersic}$ & $n$ & $r_{\rm eff}$& $a/b$&PA\\
& ($\mu$m) & (mJy) & (mJy) &   &(pc) & & (degree) \\
\hline
NICMOS & 1.60 &$<1.3$&$21\pm6$&  4$^*$& $1753\pm728$ & $0.60\pm0.03$ & $82\pm2$  \\
NICMOS & 2.22 &$5\pm2$&$25\pm6$&  4$^*$& $1091\pm584$ & $0.60\pm0.03$ & $82\pm3$ \\
CC & 8.70 &$36\pm4$ & $169\pm 10$ & $2.0\pm0.2$ & $1127\pm65$ & $0.60\pm 0.02$
& $78\pm1$\\
\hline
\end{tabular} 
\label{tab1} 

 Notes. --- Column 1: instrument. Column 2: central wavelength of the filter. Column 3:
 unresolved nuclear 
 emission. Column 4: emission of the S\'ersic component. Column 5:
 index of the S\'ersic profile. $^*$For the 1.60 and $2.22\,\mu$m model
 the index was  fixed. Column
 6: effective radius in pc of the S\'ersic component. Column 7: axis ratio
 of the S\'ersic component. Column 8: PA of the major axis of
the S\'ersic component measured East of North.

\end{minipage}
\end{table*}
 
 To constrain the unresolved emission in the near-IR and its uncertainty we use different combinations of profiles that always include a PSF profile to represent the unresolved emission. These combinations of profiles are: a PSF and a S\'ersic profile with all the parameters
  free, a PSF plus a S\'ersic profile with fixed $n=1$, a PSF plus a S\'ersic profile with fixed $n=4$, and
finally one PSF plus two  S\'ersic profiles including a
Gaussian profile\footnote{A Gaussian
  profile is a special case of the S\'ersic profile 
when $n=0.5$ and the size parameter is the FWHM instead of
$r_{\rm eff}$.}  with all the parameters left to vary freely. We find that the models that include a PSF plus two S\'ersic profiles with free parameters and a PSF plus one S\'ersic profile with n=4 produce fits with   similar $\chi^2\sim0.35$ and that the unresolved emission  measured on
  the NICMOS images did not 
change by more than $1\%$ at  
  both wavelengths. We then conclude that the unresolved
component is well constrained at both near-IR wavelengths. 
Fig.~\ref{fig2} shows the
combined NICMOS $2.22\,\mu$m image together with
the {\sc GALFIT} model image and the residual image after subtracting
the {\sc GALFIT} model from the image. 

Since the fits with different models are of similar quality, we estimate the uncertainty of the parameters in Table~2 as the standard deviation of the values given by all models. These uncertainties are larger than those estimated for a single model.

The unresolved emission detected in UGC~5101
contributes  2\% and 17\% of the continuum emission at 1.60 and $2.22\,\mu$m,
respectively, measured over the {\it HST}/NICMOS Camera 2 FOV. Our
unresolved fluxes at H and K band are lower than those reported by Scoville et
al. (2000) using the same NICMOS/HST images. This is because they only subtracted the underlying galaxy
as estimated from the adjacent pixels instead of modelling the host
galaxy emission. The H band image was also analyzed by Haan et al. (2011) using a similar decomposition technique as used here, where they found similar results for PSF contribution and S\'ersic parameters, within the uncertainties. We adopt our own values for the analysis that follows, since our estimation of PSF uncertainties follows the same methodology as that used to measure the unresolved component at K and mid-IR bands, and it considers a larger range of models than in the analysis by Haan et al. The host galaxy parameters are similar to those found in previous studies at both near-IR bands (e.g. Scoville et al. 2000, Haan et al. 2011, Medling et al. 2014).

\subsubsection{Mid-IR emission}
As done for the {\it HST}/NICMOS images, we also run {\sc GALFIT}
  on the CC $8.7\,\mu$m image to estimate  
the unresolved and integrated emission of the galaxy.  For the unresolved
  component we used the CC observation of the standard star taken
  just before the galaxy observation. We assume that the FWHM did not vary on those time scales and therefore the emission of the standard star can be used to represent the unresolved emission of UGC~5101. We  fit a PSF and
S\'ersic component with all parameters  left free to vary. The best
fit had a reduced $\chi^{2}\sim1$. The uncertainties for the
parameters are the standard deviation of the parameters derived by considering different convolution boxes.
 The lower panels of Fig.~\ref{fig2} show the 
  original CC $8.7\,\mu$m image together with the 2D {\sc GALFIT}
  model and the residual image.
After subtracting the
{\sc GALFIT} model we  obtained a residual of 
less than $1.5\%$   when compared to the flux in the original image.
Table~\ref{tab2} lists the unresolved
and S\'ersic fluxes at $8.7\,\mu$m. The unresolved component
contributes $\sim 12\%$ of the observed emission at this wavelength.  The
 uncertainties in the reported profile fluxes are the formal photometric errors
reported by {\sc GALFIT} added 
quadratically to the standard deviation ($6\%$ for both the S\'ersic
and the PSF components) of the fluxes obtained by
considering different convolution box sizes in {\sc GALFIT} 
 and to the $10\%$ flux calibration uncertainty (see Table~\ref{tab2}).

\begin{table*} 
	\begin{minipage}{1.\textwidth}
		\caption{Unresolved and integrated emission at
			$8.7\,\mu$m using three different methods.}
		\begin{tabular}{ccccc} 
			\hline
			Method & $f_{\rm unresol}$ & $r_{\rm apert}$& $f_{\rm int}$&$r_{\rm apert}$\\ [0.5ex] 
			& (mJy)        &(arcsec)  &    (mJy)  & (arcsec)\\
			\hline
			{\sc GALFIT}  & $36\pm4$  & 9.2& $205\pm11$&NA  \\
			PSF-scaling & $40\pm7$ &0.5 & $240\pm41$&9.2\\
			PSF-subtraction& $41\pm12$ & 9.2&$168\pm32$ &9.2\\
			\hline
		\end{tabular} 
		\label{tab2} \\
		Notes.--- NA=No assigned.
	\end{minipage}
\end{table*}

We  also estimated the unresolved and integrated emission at
$8.7\,\mu$m with two independent methods  routinely used in the 
  literature for ground-based 
mid-IR imaging. The first one is referred to as
{\it PSF-scaling} photometry and is
useful only in sources that are dominated by the nuclear emission
(see e.g. Mason et al. 2012). In this method, we measured the flux 
within a circular aperture of $0.5$-arcsec radius. Next,  we
scaled the PSF-star image to the centroid of the galaxy emission
and measured the flux within the same aperture around the
scaled PSF star. This method provides the maximum contribution
from an unresolved nuclear source. 
For UGC~5101 we obtained the integrated emission using a large
aperture with a  $9.2\,$arcsec radius, which is the radius at which
the curve of growth flattens.

The second method we term {\it PSF-subtraction} 
 (e.g. Radomski et al. 2002, 2003; Soifer et al. 2000; Levenson et
 al. 2009; Ramos Almeida et al. 2009, 2011; Garc\'{\i}a-Bernete et
 al. 2014).  In this method, we first
 matched the PSF-star image (see Fig.~\ref{fig1}, right panel) to the peak of the  galaxy emission, that is, at
   a $100\%$ level. Then we subtracted the scaled standard PSF-star
from the  galaxy image  at different
 percentage peak levels until we obtained a flat 
 profile in the residual image at $\lesssim 0.3$ arcsec. For the CC $8.7\,\mu$m image this was
 achieved at a subtraction level of $60\%$ of the peak intensity. We
 then measured the  
 unresolved component by integrating the emission in a
 $9.2\,$arcsec-radius aperture  on the scaled  
 PSF-star image. We  computed the integrated flux by adding the
 unresolved component to the extended component as measured by
   integrating the flux in the same aperture radius on
the galaxy image  after subtracting the scaled PSF image
(see Fig.~\ref{fig1}, right panel).

We list in Table~\ref{tab2} the unresolved fluxes at $8.7\,\mu$m using 
the {\it PSF-scaling} and {\it PSF-subtraction} methods. We 
  computed the uncertainties adding in quadrature the
photometric error\footnote{Calculated as $\sqrt{\sigma_{\rm
    back}^{2}N_{\rm pix} + \sigma_{\rm
    back}^{2}N_{\rm pix}^{2}/N_{\rm pix-ring}}$, where 
$N_{\rm pix}$ is the number of pixels inside the aperture  considered, $N_{\rm pix-ring}$ is the number of pixels inside a ring of 80 pixels width around the source used to estimate the background level and its
standard deviation $\sigma_{\rm back}$ (Reach et al. 2005). The background in our images is completely flat, and this makes the second term of the error equation almost negligible.} and the  
uncertainty in the flux calibration ($10\%$). For the {\it PSF-scaling} method we assumed a $6\%$ 
uncertainty due to time-variability, estimated from the variation in the signal from the standard stars, and a $13\%$ uncertainty due to PSF-variations. These uncertainties are derived in Mason et al. (2012) from several standard stars observed during the same night at mid-IR with T-ReCS and Michelle on the Gemini North telescope. The uncertainties assumed were added in quadrature to the errors previously described. For the {\it PSF-subtraction} method we
added in quadrature, in addition to all the above error contributions, a $7\%$ of uncertainty due to ambiguity in the best subtraction of the
unresolved 
source.
The unresolved fluxes estimated with the PSF methods are in
good agreement with the value estimated with {\sc GALFIT}.  

 In order to assess the robustness of decomposition and whether there might be a dependency with the adopted PSF, we use the acquisition image of the standard star for spectroscopy obtained 2 hours after the science image (see Table \ref{tab0}) and repeat the calculations of GALFIT decomposition of UGC~5101 with this PSF model. We find an unresolved flux $27\pm6$ mJy, which is consistent with the value derived before. We also find that the S\'ersic parameters are consistent with those previously derived and listed in Table \ref{tab1}.

We also used {\sc GALFIT} to measure the unresolved flux in the
Subaru/COMICS $Q$-band image at 17.7 $\mu$m. In this case we used a
Gaussian profile with a 
  ${\rm FWHM}\sim0.6-0.8\,$arcsec, 
which is the range FWHM of the standard star (see Asmus et
al. 2014),  to represent the unresolved component and a S\'ersic
profile for the extended component. We considered three 
S\'ersic profiles: one free, one with fixed $n=4$, and one with fixed
$n=1$. The best fit was obtained with a ${\rm FWHM}=0.65\,$arcsec for
  the unresolved component and $n=2.1\pm1.8$.
This produced an unresolved
  flux at $17.7\,\mu$m of $158\pm32$ mJy. This is lower than the
  unresolved flux calculated by Asmus et al. (2014) using
  the same data set but consistent with that in Soifer et al. (2000)
  using Keck imaging.  The 
  error in the unresolved emission is  
  the quadratic sum of  the standard deviation of the
flux measured by fitting different profiles and the formal error from
{\sc GALFIT}. 

\begin{figure*}
	\center
	\includegraphics[scale=0.3]{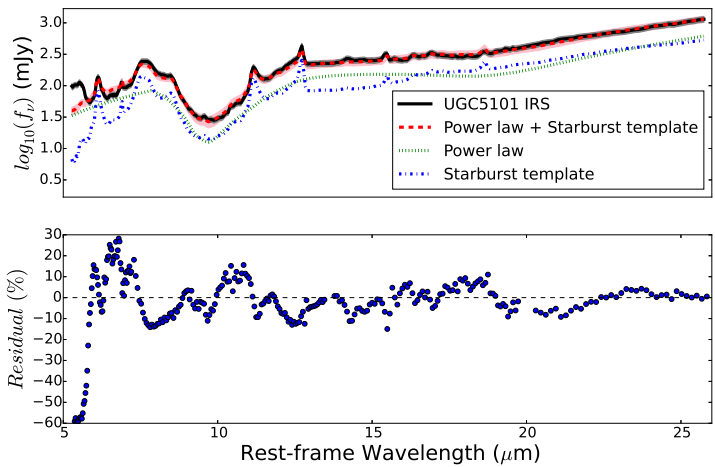}
	\caption{\label{fig4}Upper panel: spectral decomposition of the {\it Spitzer}/IRS
		$\sim 5-35\,\mu$m 
		spectrum from the central 4\,arcsec ($\sim 2.9\,$kpc) of UGC~5101. The
		black line is the observed spectrum 
		with the grey shaded region representing the $1\sigma$ uncertainty. The
		red line is the best-fit model (the pink shaded area is the 
		$1\sigma$ uncertainty of the fit), which is the sum of the
		fitted starburst template 
		(blue line) and  
		the power-law continuum component (green line). Both components
		are assumed to be obscured by
		different levels of foreground extinction using the Chiar \& Tielens
		(2006) law for the local ISM (see text and Equation~2). Bottom panel: residuals of the fit, where percentage is defined as observed spectrum minus modelled spectrum, divided by modelled spectrum.}     
\end{figure*}

\subsection{Starburst subtracted nuclear emission from the spectroscopic data}
 Both 
the {\it Spitzer}/IRS and the GTC/CC spectra show emission from 
polycyclic aromatic hydrocarbon (PAH) features (see Figs~4 and
5, and Armus et al. 2004). This indicates the presence 
of a component due to star formation activity in the regions
of $\sim 393\,$pc and $\sim 2.9\,$kpc probed by CC and IRS,
respectively. In Section~4 we will also fit the starburst-subtracted nuclear GTC/CC spectrum to
infer the properties of the obscuring material, presumably the dusty torus,
around the AGN of UGC~5101. Before we do so, we need to remove the
emission due to star formation. Spectral decomposition in the mid-IR 
has proven to be an
efficient method to disentangle the AGN emission from that due
to star formation activity in the host galaxy in luminous infrared
galaxies (LIRGs), ULIRGs, AGN, and sub-millimeter galaxies 
(e.g. Pope et al. 2008; Valiante et al. 2009;
Nardini et al. 2010; Mullaney et al. 2011; Alonso-Herrero et al. 2012; Ramos Almeida et
al. 2014b; Hern\'an-Caballero et al. 2015, Garc\'ia-Bernete et al. 2015).

For the spectral decomposition we used the model
proposed by
Pope et al. (2008). This model includes three main components:
(1)  a starburst dominated by the broad features
  due to PAH molecules up to $\sim 18\,\mu$m (Puget $\&$ Leger
1989; Allamandola et al. 1999), (2) a power-law (or warm blackbody)
continuum\footnote{Although the use of a power-law continuum in the
mid-IR (up to $\lambda \sim 
35\,\mu$m)  is not physically motivated, it represents well the AGN
heated dust continuum which usually peaks at $\lambda \sim
30-50\,\mu$m (see e.g. Alonso-Herrero et al. 2011).} that represents the obscuring material around the AGN (Genzel et al. 1998), and (3)
extinction. 
The starburst and power-law components can be affected by
different degrees of extinction, which is 
modelled as a foreground dust screen. The model is expressed as:
\begin{equation}
F_{\nu}=c_{1}\nu^{-c_2}e^{-c_3\tau_{\nu}}+c_{4}f_{\nu}({\rm PAH})e^{-c_5\tau_{\nu}},
\label{spec}
\end{equation} 
where $\tau_{\nu}$ is the extinction law,  $c_3$ and $c_5$ are the power-law component and starburst extinctions, $c_2$ is the index of the power-law,
 $f_{\nu}({\rm PAH})$ the starburst component, and $c_1$ and  $c_4$ are
 the normalizing factors.

 We used a total of 24 starburst galaxies from the literature 
 including IR-bright galaxies, LIRGs, and ULIRGs (Sturm et al. 2000;
 Brandl et al. 2006; Rieke et al. 2009; Sargsyan et al. 2011). We used  
the extinction laws of Chiar $\&$ Tielens 
(2006) for the local interstellar medium (ISM) and the Galactic Center (GC), as
they  include the absorption due to silicates around
10 and $18\,\mu$m. We  performed a $\chi^{2}$ 
minimization and fitted simultaneously  the five free
  parameters $c_{1}$, $c_{2}$, $c_{3}$, 
$c_{4}$, and $c_{5}$ for each of the starburst templates and extinction laws.

\subsubsection{Spectral decomposition of the 
Spitzer/IRS spectrum}

\begin{table}
\caption{Results from the spectral decomposition of the IRS and CC spectra.}
\begin{minipage}{0.6\textwidth}
\begin{tabular}{c|cc}
	\hline
	             Parameter              &    {\it Spitzer}/IRS    &           GTC/CC           \\
	    [0.5ex] 
\hline     
	      Power-law  index $c_2$        &       $2.0\pm0.2$       &        $2.0(fixed)$        \\
	   Power-law $\tau_{\lambda9.7}$    &       $5.1\pm0.6$       &        $6.4\pm2.0$         \\
	   Starburst $\tau_{\lambda9.7}$    &       $0.9\pm0.4$       &        $0.3\pm0.2$         \\
	$8.7\,\mu$m $f_{\rm SBsubs}$ (mJy) &        $79\pm17$        &         $53\pm35$          \\
	$12\,\mu$m $f_{\rm SBsubs}$ (mJy)  &        $62\pm18$        &         $42\pm15$          \\
	        Reduced $\chi^{2}$          &           2.3           &            3.0             \\
 \hline
\end{tabular} 
\end{minipage}
\label{tab3} 
Notes.---$f_{\rm SBsubs}$ is the flux obtained from subtracting the observed spectrum and the best-fit starburst model at $8.7\,\mu$m and $12\,\mu$m, respectively.
\end{table}

 Considering that the local ISM extinction law of Chiar \& Tielens (2006) covers a range from $1.2$ to $27\, \mu$m together with the high S/N and spectral range covered by the
{\it Spitzer}/IRS spectrum $5-35\,\mu$m compared with the GTC/CC spectrum, we started by decomposing the {\it Spitzer}/IRS spectrum
 between $5-26\,\mu$m. Then, in 
Section~3.2.2, we use the starburst galaxy obtained from the best {\it Spitzer}/IRS spectral decomposition fit to perform
  the spectral decomposition of the nuclear CC
spectrum.  The best fit for the {\it Spitzer}/IRS 
spectrum is achieved using the 
LIRG ESO~244-G012 spectrum (blue line) from Sargsyan et al. (2011) and a
power-law component with an index of $c_{2}=2.0\pm 0.2$ (green line). The index of the power-law
is similar to that found in the mid-IR for Seyfert 1s and Seyfert 2s (Armus et al. 2007; H{\"o}nig et 
al. 2010; Ramos Almeida et al. 2009, 2011). The fitted power law component in the mid-IR is consistent with the presence of a hot dust component ($T>300$ K) found by Armus et al. (2007) for this galaxy from the fitting of the infrared SED of this galaxy. 

We also found that the local ISM
Chiar \& Tielens 
(2006) extinction law produced a better fit to the data than GC one. As can be seen
from Fig.~4, the fit reproduces well  the mid-IR emission except
for the bluest part of the spectrum. This is due to the presence in
UGC~5101 of
strong absorption due to water ice under the $6.2\,\mu$m PAH feature
(see Spoon et al 2002 and Armus et al. 2004), which is not present in
the extinction law. Adding the water ice absorption
using the laboratory analogs of Gerakines et al. (1995) only improved
marginally the fit in this spectral region and  
the power-law component remained the same.

Table~\ref{tab3}
summarizes the relevant parameters of the spectral decomposition of
the IRS spectrum. The errors in the
derived parameters take into account the $10\%$ uncertainty
assumed for the {\it Spitzer}/IRS spectrum and the standard deviation of the parameters by considering different initial values for the index of the power-law $c_{2}$ (see Section 3.2.2). The starburst-subtracted fluxes
(observed, not corrected for extinction) at 
$8.7\,\mu$m and $12\,\mu$m (observed wavelengths) contribute proximately 54 and
50\% of the observed emission in the {\it Spitzer}/IRS spectrum, respectively. The starburst-subtracted
fractional contribution at $12\,\mu$m is in agreement with that
estimated by Hern\'an-Caballero et al. (2015) using the same IRS
spectrum but an independent method.
At rest-frame wavelengths
longer than approximately $15\,\mu$m the power-law component appears
to dominate
the continuum emission in the IRS spectrum, although at $\lambda_{\rm
  rest}>35\,\mu$m the starburst component has an almost equal contribution. This is also in agreement with the $30\%$ starburst contribution found by Armus et al. (2007) for UGC~5101 using different mid-IR diagnostics. 

Using the foreground extinction and silicate optical depth ratio $A_{V}/\tau_{\lambda9.7}=18$ from Roche \& Aitken (1984) for the local ISM the fitted
foreground extinction affecting the power-law component is $A_V=90\pm11\,$mag, which is necessary to reproduce the deep $9.7\,\mu$m silicate
feature. This high value is also consistent with the presence of dust
lanes in the nuclear region, as seen in the {\it HST} optical image of
this galaxy (Surace et al. 2000). 

 Finally, we note that using more typical starburst galaxy spectra, like those of Brandl et al. 2006 we obtained similar starburst-subtracted fluxes at 8.7 and 12 $\mu$m, extinctions at 9.7 $\mu$m and power-law index to those obtained with the galaxy ESO 244-G012, although the $\chi{2}$ values of the decomposition are larger.

\subsubsection{Spectral decomposition of the 
GTC/CC spectrum}
To determine the extraction type of the GTC/CC spectrum (point source
versus extended source) we measured the nuclear flux 
on the CC  $8.7\,\mu$m image
  within a $1\,$arcsec-diameter 
  circular aperture. We obtained a flux of $57\pm5$ mJy (after
  correction for point source emission). A comparison with the unresolved emission at $8.7\,\mu$m (Table~3) shows that 
for the CC slit width we can approximate the emission within the slit as a point
source. For the extraction as point source {\sc RedCan}
uses an extraction aperture that increases with wavelength to 
account for the decreasing angular resolution. It also performs a
correction to account for slit losses. We finally resampled the
  nuclear CC spectrum to 60 data points to improve the S/N ratio and
  for later fitting with {\sc CLUMPY} torus models (Section~4.2). The
    errors are computed as the quadratic sum of the $10\%$ photometric
    uncertainty
  and the rms of the four data point rebinning.

We did the spectral decomposition of the GTC/CC nuclear spectrum as
for {\it Spitzer}/IRS. However, because of the limited spectral
range of the CC spectrum, we fixed the starburst template to the one that
provided the best fit to the IRS spectrum (i.e. 
ESO~244-G012) and used the Chiar \& Tielens (2006) extinction law of
the local ISM. We found that the best-fit with a reduced $\chi^{2}=2.4$ is achieved with an index of the power-law $c_{2}=4.0$. Although this value is consistent with the range of spectral indices found by Buchanan et al. (2006) in a sample of Seyfert 1, 2, LINERs and starburst galaxies, it is larger than that obtained from the {\it Spitzer}/IRS decomposition (see Table~\ref{tab3}). Therefore, we study the behavior of the $c_{2}$ index in the {\it Spitzer}/IRS and CC decomposition by comparing the resulting parameters when we vary $c_{2}$ between 0 and 4. From this analysis we found that the power-law and starburst extinctions and, the starburst-subtracted fluxes at 8.7 and 12 $\mu$m are well constrained (see Table~4). Thus, we chose to fix the power-law index to $c_{2}=2.0$ for the spectral decomposition of the CC spectrum  for consistency. This value agrees well with that 
  derive from the spectral decomposition of the {\it Spitzer}/IRS
  spectrum and those measured in  Seyfert 1 and 2 nuclei in the mid-IR
  (H\"onig et   al. 2010; Ramos Almeida et
  al. 2009, 2011). 

 These results are not highly dependent on the starburst template library adopted. If for instance, we fit the full spectral library of Brand et al. (2006), we find an equally valid starburst decomposition ($\chi^2 = 3.0$) for NGC1222, and similar values for extinction, and starburst-subtracted fluxes.

\begin{figure}
	
	\includegraphics[scale=0.25]{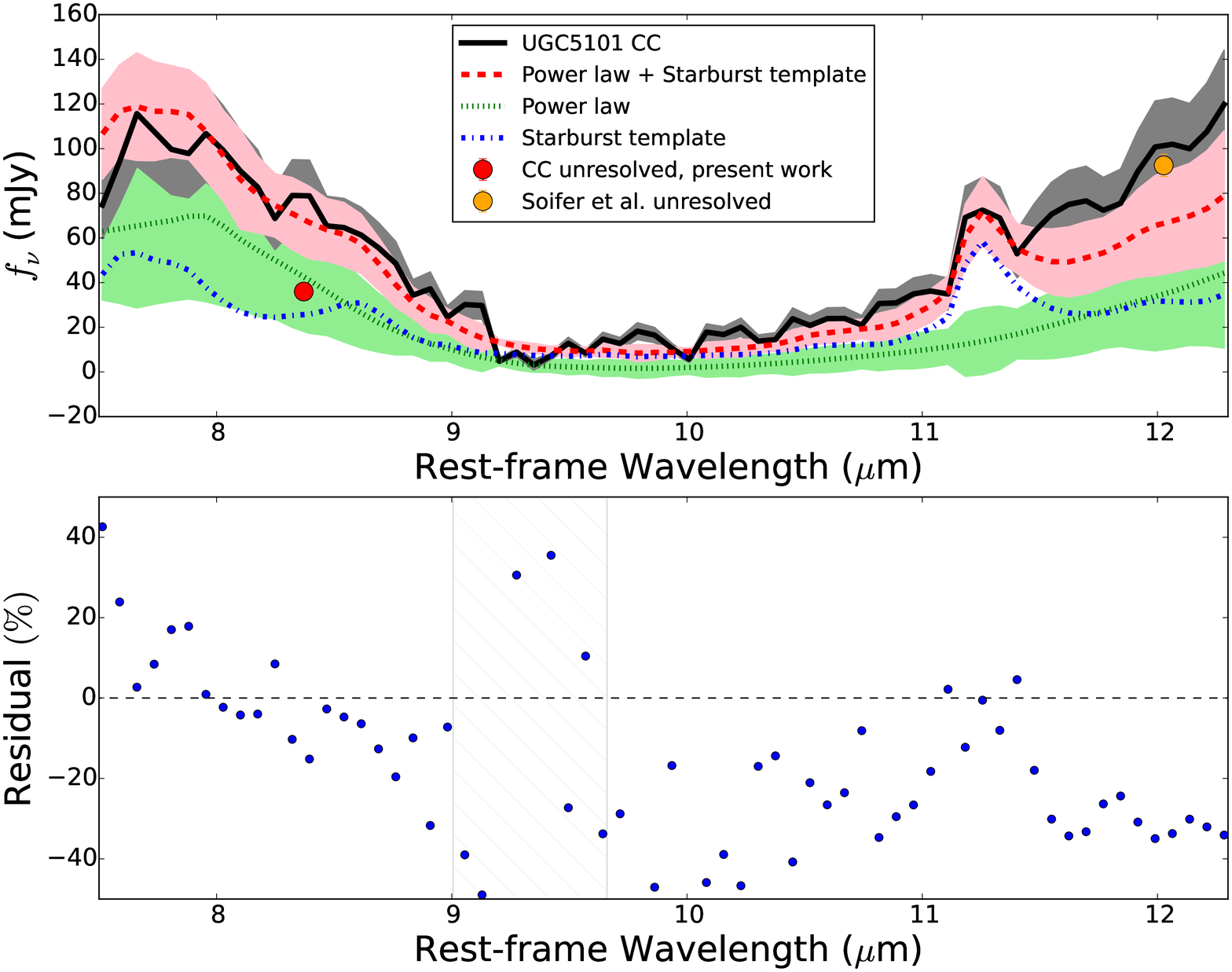}
	\caption{\label{fig5}Upper panel: spectral decomposition of the GTC/CC 
		nuclear  (inner 393\,pc) spectrum of UGC~5101. We  resampled the
		observed spectrum to
		60 data points (see text). Lines are as in
		Fig.~\ref{fig4}, where the power-law index is fixed to $c_{2}=2.0$. The filled  
		red  dot is the $8.7\,\mu$m unresolved flux estimated from the
		{\sc GALFIT} modelling in Section~3.1.2., whereas the orange filled
		dot is the $12\,\mu$m unresolved flux from Soifer et al. (2000). Bottom panel: residuals as calculated in Fig. 4. The hatched region
		represents approximately the spectral range of low atmospheric
		transmission.}     
\end{figure}

 We show the best-fit model (dashed red line and
shaded region) to the observed GTC/CC
nuclear spectrum (black line 
and grey shaded region)  in
Fig.~\ref{fig5}, with the green  and blue lines  showing the
individual contributions from the  power-law continuum and 
 the starburst components, respectively. As can be seen from this figure, 
the best-fit model reproduces well the
  data and in particular the emission of the $11.3\,\mu$m PAH
  feature. We therefore subtracted the starburst 
  component from the CC spectrum, so in the next section we
  can model the starburst-subtracted mid-IR emission.

We list in Table~\ref{tab3} the relevant parameters of the fit to
  the CC nuclear spectrum. The errors of the parameters are estimated as the standard deviation of the parameters given by different models that fix the value of $c_{2}$ between 0 and 4. The fitted foreground extinction affecting the 
  power-law continuum from the spectral decomposition of the CC
  nuclear spectrum $A_V=115\pm36\,$mag is higher than that from the IRS
  spectrum, although consistent within the uncertainties, possibly indicating that on nuclear scales the heating
  source is more embedded and the continuum
  source is extended (see below).

The starburst-subtracted fluxes, not corrected for extinction, at
  $8.7\,\mu$m and $12\,\mu$m (observed wavelengths) from the
  decomposition of the CC 
  spectrum (see Table~4) contribute approximately 67\% and 59\% to the nuclear CC 
  spectrum, respectively. Within the uncertainties these values are in good agreement with those estimated by Hern\'an-Caballero  et 
al. (2015) for the AGN component. The starburst-subtracted fluxes at $8.7\, \mu$m and $12\, \mu$m from the {\it Spitzer}/IRS and CC decomposition are consistent with each other, within the uncertainties (see Table~4). 
In general for nearby Seyfert nuclei the spectral decomposition of
IRS spectra is able to recover both the shape and flux of the
AGN component measured from high angular resolution ground-based
mid-IR spectroscopy (see e.g. Ramos Almeida et al. 2014b;
Hern\'an-Caballero et al. 2014; Garc\'{\i}a-Bernete et al. 2014). 
In the case of UGC~5101, as it is more distant than these nearby
Seyferts, it is possible that the IRS spectrum has a strong
contribution from extended dust components which are more
difficult to disentangle from
the dust heated by the AGN. 

We find, however that the
$8.7\,\mu$m starburst-subtracted nuclear flux from the decomposition of the GTC/CC nuclear
spectrum is in excellent agreement to within the  uncertainties with the unresolved emission derived from the {\sc GALFIT} model of the CC imaging data
(Section~3.1.2 and Table~\ref{tab2}). Soifer 
et al. (2000) estimated that at observed $\lambda=12.5\,\mu$m the core
component contributed approximately $50\%$ of the flux within a $4\,$ arcsec
diameter aperture ($\sim 185\pm 10\,$mJy). Within the
uncertainties, their inferred unresolved
flux is consistent with the starburst-subtracted GTC/CC spectrum at the same wavelength ($\sim91$ mJy) but not with the fitted power-law continuum (see also Fig.~5). This result shows that at this part of the spectrum the power-law is under estimated while the starburst component is well fitted.

\section{Estimating the properties of the AGN and the obscuring material}

\begin{table*}
\caption{Parameters of the {\sc CLUMPY} torus models and for the best
  fit to nuclear emission of UGC~5101.} 
\center
\begin{minipage}{0.6\textwidth}
\center
\begin{tabular}{lcccc} 
\hline\hline 
Parameter & Symbol & Interval& Median& MAP\\ 
\hline 
Torus angular width  [degrees]                     & $\sigma_{\rm torus}$ &
[15, 70] &$60^{+6}_{-7}$  & 66 \\ 
Torus radial thickness                    & $Y$                & [5, 100] & $64^{+19}_{-23}$&  87 \\
Number of clouds along an equatorial ray  & $N_{0}$          & [1, 15] & $12^{+2}_{-3}$& 13\\ 
Index of the radial density profile       & $q$                & [0, 3] & $0.8^{+0.4}_{-0.4}$& 1.0 \\
Viewing angle [degrees]                            & $i$                &
[0, 90] &$55^{+18}_{-19}$ &46 \\ 
Optical depth per single cloud            & $\tau_{V}$     & [5, 150] &$36^{+15}_{-10}$ &32 \\

\hline
\end{tabular} 
\\
 Notes.---Columns 1, 2 and 3 show the name of the {\sc CLUMPY} torus model parameters, their symbol and the fitting range. Columns 4 and 5 show the median and maximum-a-posterior values obtained from the posterior distributions. 
\end{minipage}
\label{clumpy_model}  
\end{table*}

\subsection{The {\sc CLUMPY} torus models}
\begin{figure}
\center
\vspace{0.5cm}
\includegraphics[scale=0.45]{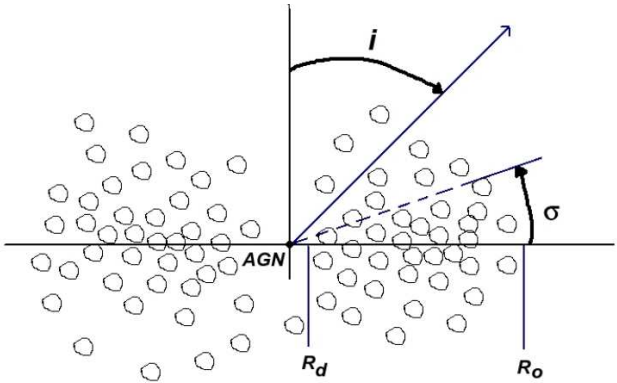} 
\vspace{0.5cm}
\caption{\label{fig6} Cartoon showing the geometry of the {\sc
      CLUMPY} dusty torus models of Nenkova et al. (2008a, b). According to this model each dusty cloud has an optical depth $\tau_{V}$, and the clouds are distributed between the inner radius $R_{d}$, determined by the dust sublimation temperature, and the outer radius $R_{o}=YR_{d}$, with $Y$ a free parameter. The angular distribution of the clouds is $\sigma$. The torus viewing angle from the observer is $i$. }
\end{figure}

The {\sc CLUMPY} model (Nenkova et al. 2002, 2008a, b) holds that the dust
surrounding the central engine of an AGN is distributed in clumps. An
illustration of this model is shown in Fig.~\ref{fig6}. 
The clumps have a radial distribution $\propto r^{-q}$  with a
radial extent 
$Y=R_{\rm o}/R_{\rm d}$, where $R_{\rm o}$ and $R_{\rm d}$ are the outer and inner
radius of the toroidal distribution, respectively. The inner radius is
defined by the dust sublimation temperature, which is assumed to
  be $T_{\rm sub}\approx1500\,$K,   and the AGN luminosity as: 
$R_{d}=0.4$ (1500\,K $T^{-1}_{\rm sub}$)$^{2.6}(L_{\rm AGN}/10^{45}$erg
s$^{-1})^{0.5}$ pc. In this model each clump has the same optical
depth ($\tau_{V}$). The number of clouds along the line of sight
(LOS) at a viewing angle $i$ is
$N_{\rm LOS}(i)=N_{0}e^{(-(i-90)^{2}/\sigma_{\rm torus}^{2})}$, where
$\sigma_{\rm torus}$ is the
angular width occupied by the cloud distribution and  $N_{0}$ is the
average number of clouds along a radial equatorial ray.  The {\sc
    CLUMPY} models have been proven to reproduce the nuclear IR
  emission of Seyfert galaxies and quasars (e.g. Ramos Almeida et
  al. 2009, 2011, 2014a,b; Nikutta et al. 2009; Mor et al. 2009;
  Alonso-Herrero  et al. 2011, 2013; Lira et al. 2013; Mori et al. 2014;
  Ruschel-Dutra et al. 2015; Ichikawa et al. 2015;
  Garc\'{\i}a-Bernete et al. 2014).

In this work we use an updated version of {\sc bayesclumpy}, a
Bayesian tool specifically designed  to fit  the {\sc
    CLUMPY} models to observed SEDs  and mid-IR spectroscopy
(Asensio Ramos $\&$ Ramos Almeida 2009). Within {\sc
    bayesclumpy} there are two additional
parameters not listed in Table \ref{clumpy_model} that can be fitted
or fixed. The first one is the scaling factor called {\it shift}, which accounts for
the vertical 
displacement needed to match the fluxes of a given model to an
observed SED and/or spectrum. This is proportional to
the AGN bolometric 
luminosity. The second one is the foreground extinction $A_{V}$ due to the host galaxy which is different from that
produced by the torus along the LOS. We use the Chiar \& Tielens
  (2006) extinction 
  law derived for the local ISM, as it provided a better fit to
  the IRS spectrum (Section~3.2.1). Foreground extinction has been shown to be 
  required to fit the nuclear IR emission of Seyfert nuclei with deep
  silicate features (see 
  e.g. Alonso-Herrero et al. 2011; Gonz\'alez-Mart\'{\i}n et
  al. 2013), as is the case for UGC~5101. We assume a uniform prior
distribution for both of 
  them, with the foreground extinction being in the range 
$A_{V}({\rm foreground)}=0-40\,$mag. {\sc bayesclumpy} can also incorporate 
the direct AGN emission to model the data, although in principle there
is no clear detection of broad emission lines in UGC~5101 (see next section).

Under the assumption that the torus models are valid simultaneously
for photometric and spectroscopic data and that the noise in all
observed data is uncorrelated, the Bayesian approach is insensitive to
the fact that there are many more spectroscopic points than
photometric ones. This is justified by the fact that the method
samples the full posterior distributions and obtains marginalized
posteriors for each parameter distribution (Asensio Ramos $\&$ Ramos
Almeida 2009).  The results of the fitting process are
the marginal posterior distributions for the six free parameters that
describe the {\sc CLUMPY} dusty torus model plus the vertical {\it
  shift} and foreground extinction. 

\begin{figure*}
\center

\hspace{-.8cm}\includegraphics[scale=0.31]{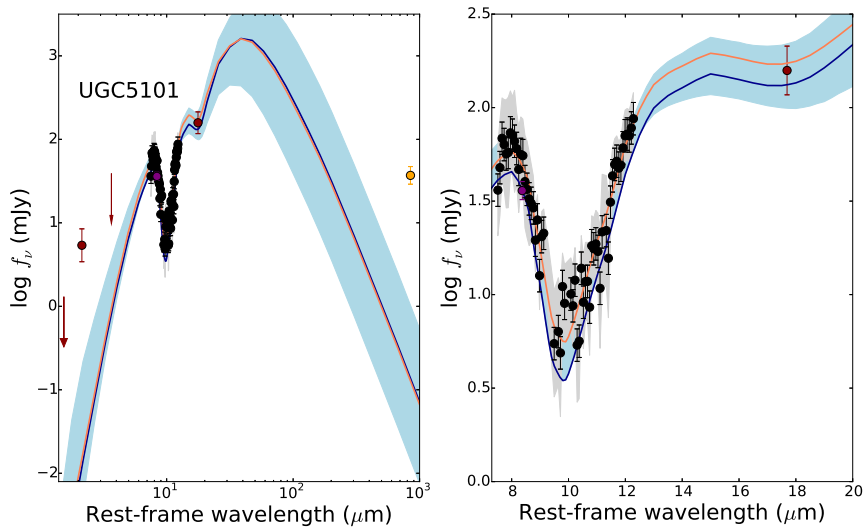}

\caption{\label{BCSED} Left: observed unresolved nuclear SED
  (red dots) and mid-IR spectroscopy  after subtracting the
starburst contribution from the  
CC nuclear spectrum (black dots). The orange dot is the nuclear $880\,\mu$m continuum
flux from Wilson et al. (2008). We did not use this data point for the
{\sc BayesClumpy} fit
(see text). The blue line
  and shaded region represent the best {\sc CLUMPY} torus model and the
  range of models   within the 68~per cent uncertainty in the best fitted
  parameters, whereas the solid orange line is the MAP model. 
Right: enlarged view of the best-fit {\sc CLUMPY}
    torus models (as in the left panel) around the $9.7\, \mu$m and $18\, \mu$m silicate features.}      
\end{figure*}

\subsection{Results from the fit to the unresolved SED plus the starburst-subtracted spectroscopy}
We modelled the spectral energy distribution (SED) and
spectroscopy of 
UGC~5101 using  the near- and mid-IR unresolved fluxes plus the starburst-subtracted GTC/CC spectrum.  
  We used as an upper limit the nuclear magnitude in the
  $L$-band from Imanishi et al. (2014). Both the observed unresolved SED and GTC/CC nuclear starburst-subtracted spectrum of UGC~5101 are
plotted in Fig.~\ref{BCSED}.

After fitting the unresolved emission and the starburst-subtracted nuclear spectrum, we obtained the
marginal posterior distributions 
for the six free parameters that describe the {\sc CLUMPY} dusty torus
model (Table~5) plus the vertical {\it shift} and foreground extinction $A_{V}$. If the
observational data contain sufficient information for the fit, then
the resulting marginal posterior distributions are centred or show
trends at certain values within the considered intervals. From
  these posterior distributions we also derived other torus parameters
  listed in Table~6.

\begin{figure*}
\center
\includegraphics[scale=0.3]{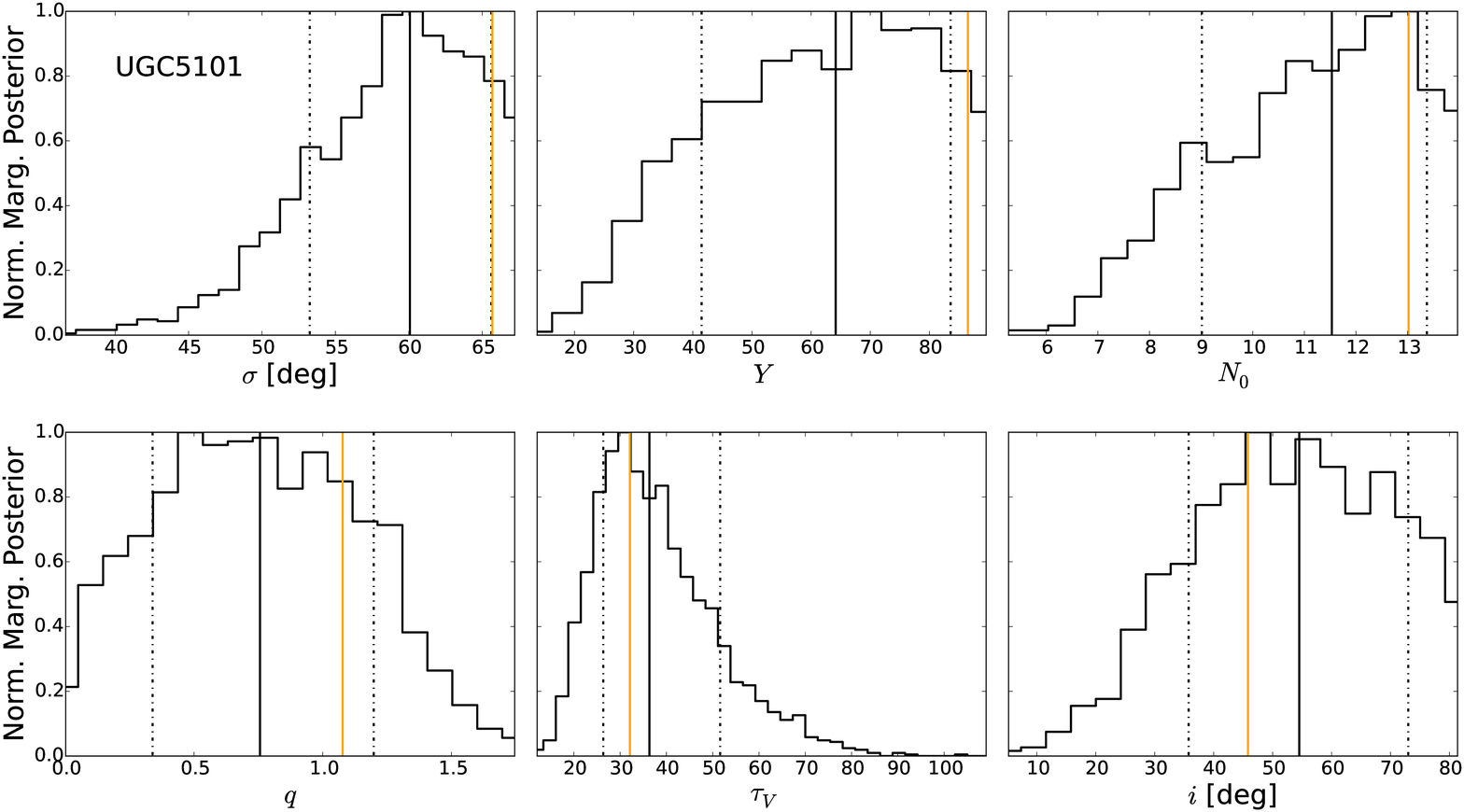}
\caption{\label{BCposteriors}Marginal posterior distributions of the
    fitted {\sc CLUMPY} torus model parameters to the nuclear emission
    of UGC~5101. The dotted lines
    represent the $1\sigma$ 
    intervals, whereas the solid black and orange lines are the MAP and
  median values of the distributions, respectively. }
\end{figure*}

The posterior distributions of the fitted torus model parameters 
show that they are well constrained (see Fig.~\ref{BCposteriors} and Table~5).
Both the high value of the angular width of the torus 
$\sigma_{\rm torus}\sim60\,$degree and number 
of clouds along the equatorial direction $N_{0}\sim12$ produce  
high apparent and geometrical covering factors $f(i)\sim0.94$ and $f_{2}\sim0.94$, which are defined as (see Alonso-Herrero et al. 2011 for details):
\begin{equation}
f(i)=\frac{L_{torus}(i)}{L_{bol}^{model}(AGN)}	
\end{equation}
and,

\begin{equation}
	f_{2}=1-\int_{0}^{\pi/2}P_{esc}(\beta)cos(\beta)d\beta.
\end{equation}
 
  These high covering factors are consistent with the derived low escape probability for an AGN produced photon $P_{esc}$ (see Table~6), which might explain the
lack of strong AGN signatures in the optical spectrum of UGC~5101. The fitted angular size and number of clouds along the equatorial direction of the UGC~5101 torus are consistent with the values found for Seyfert 2 nuclei, whereas the torus size of UGC~5101 ($\sim35$ pc) is larger than the typical sizes derived for Seyfert 1 and 2 nuclei using the same modeling techniques (Ramos Almeida et al. 2011, Alonso-Herrero et al. 2011, Ichikawa et al. 2015).
	

The high value of the foreground extinction, $A_V=32^{+4}_{-7}\,$mag, is
necessary to reproduce the deep $9.7\,\mu$m silicate feature observed in the
nuclear region. However, compared with the foreground extinctions derived from the {\it Spitzer}/IRS and CC spectral decompositions this value is nearly 3 times lower. This difference can be explained considering that the {\sc CLUMPY} models include the geometrical properties of the torus and, on this model, the foreground extinction is not the only contributor to the observed deep $9.7\,\mu$m silicate feature. The best-fit models also reproduce well the $17.7\, \mu$m unresolved flux around the $18\, \mu$m silicate feature.
All these values 
are also  consistent
with the parameters, including the high covering factors, derived in
other LIRGs and ULIRGs 
  hosting deeply embedded AGN, e.g.
Arp~299A (Alonso-Herrero et al. 2013), NGC~6240 (Mori et al. 2014), and Mrk~1066 (Ramos Almeida et al. 2014b).

We translated the fitted torus model parameters into two model
spectra. The first one corresponds to the maximum-a-posterior (MAP)
values that represent the best fit to the data. The second one is
produced with the median values of the probability distributions of
the model parameters. We plot these best-fit {\sc CLUMPY} models in
Fig.~\ref{BCSED}  
together with the both unresolved SED and starburst-subtracted GTC/CC spectrum. We also show in this figure the 
  $880\,\mu$m continuum flux of $37\pm9\,$mJy measured with a $1\,{\rm
  arcsec} \times 0.7\,{\rm arcsec}$ beam (Wilson et al. 2008),
although we did not use it for the fit. Clearly, the torus emission is
well below this flux, indicating that in the far-IR the nuclear
emission is dominated by star formation activity, as already hinted by
the spectral decomposition of the IRS spectrum
(Section~3.2.1). Additionally, the $880\,\mu$m flux may have a significant
contribution from non-thermal or radio free-free emission (see
discussion by Wilson
et al. 2008).

As can be seen from Fig.~\ref{BCSED}, the best-fit {\sc CLUMPY}
models reproduce well the shape of the silicate feature and the mid-IR
unresolved emission. The $2.2\, \mu$m is not well fitted by the {\sc CLUMPY} models.  One possibility is that the near-IR unresolved emission is still contaminated by stellar emission from recent
star formation in the nuclear region of the galaxy. However, in this case, the SED of the unresolved emission should be bluer than the observed nuclear $H-K=1.9$ color derived in this work, and at least 2 orders of magnitude more luminous than the AGN. Another explanation, more consistent with the nuclear $H-K$ color, could be the presence of very hot dust clouds of graphite ($T\sim1000$ K) (see for example, Mor \& Trakhtenbrot 2011). In fact, Alonso-Herrero et al. (2011) found a slight excess at near-IR emission in the modelling of the SED+spectroscopy of some Seyfert 1s and, Mor et al. (2009) also consider a hot dust component for modelling the SED+spectroscopy of a sample of PG quasars.  

Thus, may be in UGC 5101, a combination of {\sc CLUMPY} 
torus emission+cold foreground absorbing dust screen might not be the
appropriate 
model for active nuclei with very deep silicate features
(e.g. Alonso-Herrero et al. 2011; Mori 
et al. 2014; Merlo et al. 2014) and spherically symmetric smooth models
may be more 
appropriate for deeply embedded objects (see Levenson et al. 2007;
Alonso-Herrero et al. 2013).

\begin{table}
\caption{Derived properties from the posterior distributions of the fitted {\sc CLUMPY} torus parameters.} 

\center
\begin{tabular}{lc} 
\hline\hline 
Parameter &  Median\\ 
\hline 
Extinction-corrected $L_{\rm bol}$ [erg s$^{-1}$] & $(1.9^{+1.2}_{-0.7})\times10^{45}$     \\
Escape probability $P_{\rm esc}$ & $0.0005^{+0.0040}_{-0.0005}$ \\
Geometrical covering factor $f_{2}$&  $0.94^{+0.04}_{-0.07}$\\
Apparent covering factor $f(i)$ & $0.94^{+0.20}_{-0.15}$\\
Foreground extinction $A_V$ [mag] & $32^{+4}_{-7}$   \\

\hline
\end{tabular} 
\label{derived_par}
 Notes.--- See equation 2 and 3 for the definition of the covering factors and escape probability. 
\end{table}

\subsection{The AGN luminosity}

From the scaling of the fitted {\sc CLUMPY} models to the data we
derived an AGN
bolometric luminosity of $L_{\rm bol}=(1.9^{+1.2}_{-0.7})\times10^{45}$ erg s$^{-1}$. Using the
hard X-ray vs. 12 $\mu$m correlation observed for local Seyfert
galaxies (Gandhi et al. 2009; Levenson et al. 2009; Asmus et al. 2011) 
and the emission
contribution by the starburst-subtracted nuclear component at 12$\mu$m from the GTC/CC
spectral decomposition after correcting for the derived foreground extinction (see Table~4), we predict an intrinsic
hard X-ray luminosity  $L_{\rm 2-10 
  keV}=(4.1\pm0.7)\times10^{43}$ erg s$^{-1}$. Applying a bolometric
correction  $[10-20]\times L_{\rm 2-10 keV}$ (Marconi et al. 2004) we found a
bolometric luminosity between $4.1\times10^{44}\le L_{\rm bol}\le
8.2\times10^{44}$ erg s$^{-1}$. This range of AGN bolometric luminosities is lower than our estimate from the {\sc CLUMPY} torus model fit to the unresolved nuclear IR emission and the starburst-subtracted nuclear CC spectrum of UGC~5101.

Gonz\'alez-Mart\'in et
al. (2009)  on the other hand, estimated an X-ray
luminosity $L_{\rm 2-10 
  keV}=7\times10^{43}$ erg s$^{-1}$ after applying a Compton-thick
correction to the observed $2-10\,$keV emission. If we consider the above
bolometric correction, then the AGN bolometric 
luminosity is consistent with the value derived from fitting the {\sc CLUMPY}
torus models to the unresolved IR emission and starburst-subtracted nuclear CC spectrum with that predict from the
hard X-ray vs. 12 $\mu$m correlation. 

The inferred AGN bolometric luminosity from the {\sc CLUMPY} torus
  modelling accounts approximately for
  $\sim 
  56\%$ of the IR luminosity of the system, which is determined using the fluxes in all four {\sc IRAS}  bands (12, 25, 60 and 100 $\mu$m) and the flux-luminosity correlation in Sanders 
\& Mirabel (1996). Considering the uncertainties in the AGN bolometric luminosity (see Table~6) this value is
  consistent 
  with other estimates in the literature. For instance, Lonsdale et
  al. (2003) 
  using high angular resolution radio 
  observations estimated a 30\% AGN contribution in UGC~5101. Veilleux
  et al. (2009) derived a similar value ($35\%$) using a number of spectral
  mid-IR diagnostics using the {\it Spitzer}/IRS spectroscopy.

In summary, despite the fact that the {\sc CLUMPY} models do not fit
simultaneously all the near- and mid-IR  unresolved emission, and starburst-subtracted nuclear CC spectrum of UGC~5101,
the derived AGN bolometric luminosity is consistent with estimates
using other methods. In our modelling
the AGN bolometric luminosity is only a function of the scaling factor
of the models to the data, which is driven mostly by the fluxes near
the peak of the torus emission. In the case of UGC~5101, the peak of
the best-fit {\sc CLUMPY} torus model is 
at $\lambda_{\rm rest}\sim 40-50\,\mu$m (see Fig.~\ref{BCSED}, left panel) and
the $Q$-band data point 
(observed $17.7\,\mu$m) is the closest to 
that peak. Therefore even for deeply embedded AGN, the
modelling of the unresolved nuclear emission and the starburst-subtracted spectroscopy with the {\sc CLUMPY} torus
models can still be useful for deriving the AGN bolometric luminosity,
as this quantity is driven by the scaling of the model and is less
subject to
the detailed modelling of the near and mid-IR SED and the silicate feature.

\section{Summary and conclusions} 
We have presented new high angular resolution ($0.3-0.4\,$arcsec)
 $8.7\,\mu$m mid-IR imaging and $7.5-13\,\mu$m
  spectroscopy of the nuclear region of the ULIRG UGC~5101 using CC on
  the GTC. We also
  analyzed archival {\it HST}/NICMOS near-IR and Subaru/COMICS $Q$-band
  imaging and {\it Spitzer}/IRS 
  spectroscopy. 
The presence of  PAH emission  in the nuclear
(central $\sim 393$\,pc) CC spectrum  
indicates that some of the dust in central region is being  heated by
the UV radiation of relatively young stars. Therefore, to study
the physical and geometrical properties of the obscuring material
surrounding the AGN  of UGC~5101,  it was necessary to disentangle the IR 
emission presumably due to dust heated by the AGN from that due to
nuclear star formation and/or stellar emission.   

By using {\sc GALFIT} we were able to separate the unresolved
nuclear emission from 
the extended host galaxy emission using the near-IR (1.60 and
$2.22\,\mu$m) and mid-IR (8.7 
and $17.7\,\mu$m) imaging data. We also calculated the unresolved
emission at $8.7\,\mu$m with  other methods, namely {\it
  PSF-subtraction} and {\it 
  PSF-scaling} and found that the resulting fluxes were similar
to those derived with {\sc GALFIT}. In addition, we found that
the S\'ersic  
parameters, which represent the extended emission, derived at
near- and mid-IR wavelengths are consistent with each other.  
 
For the mid-IR spectroscopy we used a spectral decomposition
  method to estimate the emission of the obscuring material around the AGN, which we represented as an
  extinguished power-law continuum. We also allowed both components to have different foreground extinctions. From the spectral decomposition of the IRS data we found an index of the power-law ($c_{2}=2.0\pm0.2$) consistent with the values observed for Seyfert 1 and Seyfert 2 nuclei (Ramos Almeida et al. 2011). Given the limited spectral  range of the CC spectrum for the decomposition we fixed the index of the power-law and starburst template to those derived for the IRS data.
Within the uncertainties the starburst-subtracted nuclear fluxes at 8.7 and
 $12\,\mu$m CC continuum component are consistent with those derived from
 the spectral decomposition of the IRS spectrum. In addition, the starburst-subtracted nuclear flux at $8.7\,\mu$m
 is also consistent within the uncertainties with the value derived with the {\sc GALFIT} modelling of the CC
 imaging data. We also estimated that in the central 393\,pc of
   UGC~5101 the power-law component at $8.7\,\mu$m and $12\,\mu$m
   (observed wavelengths) contributes $67\%$ and $59\%$ 
 of the nuclear emission, respectively.

Before fitting the unresolved SED of UGC~5101 with the
  {\sc CLUMPY} torus models of Nenkova et al. (2008a, b), we subtracted the
  fitted starburst component from the GTC/CC nuclear observed spectrum. 
   We found that {\sc CLUMPY} torus models with a high covering factor
   and a    high foreground 
    extinction ($A_V\sim 32\,$mag) fit well
  the silicate feature and unresolved mid-IR emission. The high value of the foreground extinction is
similar to results for other deeply 
  embedded AGN in local LIRGs and ULIRGs (see e.g. Alonso-Herrero et
  al. 2013 and Mori et al. 2014).  However, the observed
unresolved near-IR 
emission of UGC~5101 is well above the fitted {\sc CLUMPY} models. This could be due to contamination by nuclear stellar emission even at the angular
  resolution of   $0.15\,$arcsec of the NICMOS data or more likely due to the presence of hot dust clouds. We also showed that
  the observed $880\,\mu$m nuclear flux is much higher than the
  extrapolation to the far-IR of the fitted torus models. Again, this
  indicates that there is strong nuclear star formation, as also
  indicated by the presence of PAH emission in the CC spectrum, and
  this component also dominates the far-IR nuclear emission of UGC~5101.

Even though the {\sc CLUMPY} torus models are not able to reproduce
simultaneously the nuclear unresolved near and mid-IR emission, and the starburst-subtracted GTC/CC nuclear spectrum of UGC~5101, the derived 
AGN bolometric luminosity ($L_{\rm bol} \sim 1.9\times10^{45}\,$erg s$^{-1}$)  is
consistent with that estimated from hard X-rays (Gonz\'alez-Mart\'in
et al. 2009) but higher than that from the $2-10$keV vs. $12\, \mu$m correlation (Gandhi et al. 2009; Levenson et al. 2009; Asmus et al. 2011).

\section*{Acknowledgement}

We thank the referee for a thorough report that has improved the paper significantly. This work has been partly supported by Mexican CONACyT
under research grant 
CB-2011-01-167291. MM-P acknowledges support  by the CONACyT PhD
fellowship program. AA-H and AH-C acknowledge 
financial support from the Spanish Plan
Nacional de Astronom\'{\i}a y Astrof\'{\i}sica under grant
AYA2012-31447, which is party funded by the FEDER programme, and the Universidad de
Cantabria through the Augusto G. Linares programme.
CRA is supported by a Marie Curie Intra European
Fellowship within the 7th European Community Framework Programme
(PIEF-GA-2012-327934). IG-B ackowledges financial support from the
Instituto de Astrof\'{\i}sica  
de Canarias through Fundaci\'on La Caixa and from the Spanish Ministry
of Science and Innovation (MICINN) through project 
PN AYA2013-47742-c4-2-P (Estallidos).

This work is based on observations made with the GTC, installed
in the Spanish Observatorio del Roque de los Muchachos of the
Instituto de Astrof\'{\i}sica de Canarias, in the island of La Palma. It is
also based partly on observations obtained with the Spitzer Space
Observatory, which is operated by JPL, Caltech, under NASA contract
1407. This research has made use of the NASA/IPAC Extragalactic
Database (NED) which is operated by JPL, Caltech, under
contract with the National Aeronautics and Space Administration.
The CASSIS is a product of the Infrared Science Center at Cornell
University, supported by NASA and JPL.
Based on observations made with the NASA/ESA Hubble Space Telescope,
obtained from the data archive at the Space Telescope Science
Institute. STScI is operated by the Association of Universities for
Research in Astronomy, Inc. under NASA contract NAS 5-26555.

\newpage

%

%
\end{document}